\documentclass[aps,prd,twocolumn,showpacs,amsmath,amssymb,nofootinbib]{revtex4}
\usepackage{dcolumn}
\usepackage{bm}
\usepackage{amsmath}     
\usepackage{amsfonts}
\usepackage{amssymb}
\usepackage{natbib}
\usepackage{latexsym}
\usepackage{exscale}
\usepackage[dvips]{graphicx,epsfig}

\def\be{\begin{equation}}
\def\ee{\end{equation}}
\def\bea{\begin{eqnarray}}
\def\eea{\end{eqnarray}}

\begin{document}

\title{Non-Gaussianity and the CMB Bispectrum: confusion between
         primordial and primary-Lensing-Rees Sciama contribution?}
\author{Anna Mangilli$^{1,2}$, Licia Verde$^{1,2,3}$}
\affiliation{$^1$ Institute of Space Sciences (IEEC-CSIC), Fac. Ciencies,
Campus UAB, Bellaterra, Spain\\
$^2$ ICC-UB (Instituto de Ciencias del Cosmos at Universitad de Barcelona), Marti i Franques 1, Barcelona, Spain.\\   
$^3$ ICREA (Institucio' Catalana de Recerca i Estudis Avancats)}
\begin{abstract}
We revisit  the predictions for the expected  Cosmic Microwave Background bispectrum signal from the  cross-correlation of  the primary-
lensing-Rees-Sciama signal; we point out that it can be a significant contaminant  to the bispectrum signal from primordial non-Gaussianity of the local type. This non-Gaussianity, usually parameterized by the non-Gaussian parameter $f_{NL}$, arises for example in multi-field inflation. In particular both signals are frequency-independent,  and are maximized for  nearly squeezed configurations. While their detailed scale-dependence and harmonic imprints are different for generic bispectrum shapes,  we show that, if not included in the modeling, the  
primary-lensing-Rees-Sciama contribution yields an effective $f_{NL}$ of $10$ when using a bispectrum estimator optimized for local non-Gaussianity. Considering that expected 1-$\sigma$ errors on $f_{NL}$ are $< 10$ from  forthcoming experiments, we conclude that the contribution from this signal must be included in future constraints on $f_{NL}$ from the Cosmic Microwave Background bispectrum.
 
\end{abstract}
\maketitle

\section{Introduction}
The increased sensitivity of the forthcoming Cosmic Microwave Background (CMB) experiments will open the possibility to detect higher-order correlations in the CMB temperature fluctuations beyond the power spectrum. This means that it would be possible to   study in detail eventual deviations from Gaussian initial conditions and thus gain an unique insight into the physics of the early universe (see e.g.\cite{Bartoloetal04} for a review). 
Since gravitationally-induced non-linearities  at last scattering are much smaller than in the late-time universe, the CMB is expected to be the best probe of the primordial non-Gaussianity (e.g., \cite{VWHK00,luo-94}).

Moreover, even in absence of these primordial deviations, measuring the  CMB three-point correlation function or, equivalently its Fourier analogue, the angular bispectrum, would be very useful to trace the imprint of the non-linear growth of structures on secondary (i.e. late-time)
anisotropies and would open a new window into the understanding of the evolution and growth of structures. The expected bispectrum signature of secondary 
CMB anisotropies has  
been studied in e.g.,
\cite{Goldberg&Spergel-I, Goldberg&Spergel-II, KomatsuPhD, licia, CoorayHu2001,Castro04,serra-2008,arXiv-Hanson}. In addition, non-linear physics happening between the end of inflation and the last scattering surface may leave some imprints in the CMB bispectrum (e.g., \cite{Bartoloetal04, ref+1,ref+2,ref+3,VerdeMatarrese} for local-type non-Gaussianity, which is relevant here).

To clarify the use of our nomenclature, by {\it primary} non-Gaussianity (or primary bispectrum) we refer to the combined effect of primordial non-gaussianity and of the physics happening between the end of inflation and the last scattering surface. This name is chosen accordingly to the CMB nomenclature of  ``primary anisotropies" which are related to the primordial ones but are further processed. In the literature what we call ``primary" non-Gaussianity is  often loosely referred to as ``primordial". Analogously, we use  {\it secondary} non-gaussianity (or secondary bispectrum) to refer to late-time physics, as it is usually done for secondaries CMB anisotropies. Note that  the Integrated Sachs Wolfe effect is thus considered a secondary anisotropy.

After galactic foregrounds, point sources and the Sunyaev-Zeldovich  
signature from clusters are expected to be the dominant source of  
non-Gaussianity in CMB data.  At $\ell <1500$ the statistical properties of these  
two contributions is expected to be quite similar: both are expected  
to behave as an extra Poisson contribution (see \cite{KomatsuSpergel01}
for treatment). In addition, both signals have a well known frequency  
dependence that can be used to clean CMB maps from this contaminating  
signal. 

In \cite{licia} it was shown  that the leading contribution to the CMB  secondary bispectrum with a  
blackbody frequency dependence is that of the primary-lensing-Rees Sciama correlation, where by Rees-Sciama  we mean the combination of the linear effect (Integrated Sachs Wolfe) and of the non-linear one.  Note that in the original paper it was called primordial-lensing-Rees Sciama, because of the loose nomenclature convention  explained above. Ref \cite{licia} also computed the  expected signal-to-noise for  this effect and found that experiments  
such a Planck \footnote{www.rssd.esa.int/Planck} or  ACT  \footnote{http://www.physics.princeton.edu/act/} should be able to obtain a high statistical  
significance detection. This was then independently confirmed by \cite{Giovi2003,Giovi2005}.
Note that there are  sources of non gaussianity  that are not strictly primordial (inflationary) but they arise between the end of inflation and the last scattering surface. These may well dominate over the inflationary contribution but they are mostly of equilateral type (e.g., \cite{ref+3,BartoloRiotto,Pitrouetal}; as we will see below, here we concentrate on  the local (squeezed) type.

Sparked by a recent study claiming a more than 95\% confidence limit evidence  
\cite{Yadav08} of local non-gaussianity, but   see \cite{SpergelWMAP3, CreminelliWMAP3, KomatsuWMAP5, Smithwmap5},   the subject of  primordial non-Gaussianity has received renewed attention.  We are  
thus motivated to revisit the predictions for the expected bispectrum  
signal from the primary-
lensing-Rees Sciama correlation in light  
of new developments since the year 2000: a much better-determined  
fiducial cosmological model with a lower $\sigma_8$, improved  
description of linear and non-linear evolution of  clustering in the  
presence of dark energy, optimized estimators for primordial non-Gaussianity via the bispectrum signal and the tantalizing hint of a  
possible detection.
Instead of  concentrating on the usefulness of the primary-lensing-Rees Sciama bispectrum  in constraining dark energy as done so far in the literature, we will   explore whether it could  be confused with the primary (or primordial) non-Gaussian signal and examine  
possible ways to separate the two.

The rest of the paper is organized as follows: in Sec. \ref{theory} we review the  
basic of the primary CMB bispectrum and of the primary-
lensing- Rees Sciama one. In Sec. \ref{numerical} we present numerical results for the expected signal-to-noise, the dependence of the signal on bispectrum shape and we  quantify  
the dependence of the secondary 
bispectrum signal on different descriptions for 
non-linear evolution of clustering.  In Sec. \ref{shape} we explore a possible  
confusion between the two bispectra and prospects for separating the  
signals. Finally we conclude in Sec. \ref{end}.  Useful formulae are  
reported in the Appendix.

\section{The  Primary and the secondary Lensing-Rees-Sciama CMB bispectrum}\label{theory} 

In this section we review the necessary background and the basic description of the
primary CMB angular bispectrum and the
secondary one arising from the cross correlation among the primary-lensing-Rees Sciama (L-RS) contributions. 
For the primordial non-Gaussianity we will consider the so-called {\it local} type characterized by a
momentum-independent non-Gaussian parameter -$f_{NL}$- and  refer to \cite{KomatsuPhD, KomatsuSpergel01, Bartoloetal04}.  This is the workhorse model for testing deviations from gaussianity both in CMB data and in large-scale structure data. In reviewing the secondary L-RS bispectrum we mainly follow \cite{Goldberg&Spergel-II,licia,Giovi2003}.

 \subsection{Primordial Non-Gaussanity}

In order to study higher-order statistics and model 
small deviations from Gaussianity, one can define the 3-point correlation function of Bardeen's curvature perturbations \footnote{Note that the Newtonian gravitational potential $\phi$ has the opposite sign of Bardeen's curvature perturbations: $\Phi=-\phi$} in momentum space, $\Phi(\bold k)$, as:
\be\label{eq1}
\langle \Phi(\bold {k}_1)  \Phi(\bold {k}_2)  \Phi(\bold {k}_3) \rangle\!=\!(2 \pi)^3 \delta^3(\bold {k}_1+\bold {k}_2+\bold {k}_3) F(k_1,k_2,k_3),
\ee
where the function $F(k_1,k_2,k_3)$ describes the correlation among the modes and depends on the shape of the $(\bold {k}_1,\bold {k}_2,\bold {k}_3)$ triangle in momentum space.
Different models make different predictions for the function $F$, depending on the mechanism of production of such a correlation \cite{liguori-06,FergussonShellard08}.

There are two main, physically-motivated classes \cite{creminelli-local}:

{\it i)} Local form (Squeezed configurations).\\
 This non-Gaussianity arises from the non-linear relation between the light scalar field (different from the inflaton) driving the perturbations and the observed $\Phi$.  
 In the weak non-linear coupling case the non-Gaussianity can be parametrized in real space as in Eq.~(\ref{fnl}) with $f_{NL}$ quantifying the ``level" of non-linearity. Being {\it local} in position space, it couples Fourier modes  far outside the horizon. The signal is maximal for squeezed triangle configurations with the coupling of one large-scale mode with two small scale modes.   Examples of this class of models can be the curvaton scenario \cite{lyth-curvaton} or the Ekpyrotic model \cite{koyama-07, buchbinder-2007}. 

{\it ii)} Non-local form (Equilateral configurations).\\
In this case the correlation among modes is due to higher derivative operators for single field models with non-minimal coupled Lagrangian. Such a correlation is strong for modes with comparable wavelength so that the signal is maximal for equilateral configurations. Examples of models of this kind are the ghost inflation \cite{ghost} and the Dirac-Born-Infeld (DBI) \cite{DBI} models among others.

In the following we will focus on the the first class of models  where non-Gaussianity can be parametrized as 
 \be \label{fnl}
 \Phi(\bold x) = \Phi_L(\bold x) + f_{NL} (\Phi^2_L(\bold x) - \left< \Phi^2_L(\bold x) \right>),
 \ee
where $ \Phi(\bold x) $ denotes the Bardeen potential, $\Phi_L(\bold x)$ denotes the linear Gaussian part of the perturbation and the $f_{NL}$ is a merely multiplicative constant that quantify the level of non-Gaussianity.

The $ \Phi (\bold k)$-field bispectrum will thus  have contributions of  the form:
\bea\label{phi-bisp}
&& \left <  \Phi_L (\bold{k}_1)  \Phi_L (\bold{k}_2)  \Phi_{NL} (\bold{k}_3) \right>= \\
&=&  2 f_{NL}\,(2 \pi)^3 P_\Phi (k_1) P_\Phi (k_2) \delta^3 (\bold{k}_1+\bold{k}_2 + \bold{k}_3) +cyc.,\nonumber
\eea
(i.e.  the function $F$ of Eq.~\ref{eq1} is $F(k_1,k_2,k_3)=2 f_{NL}P(k_1)P(k_2) +cyc.$)
 where we have used  the definition of the Bardeen's potential linear power spectrum\footnote{Recall that $\left< \Phi^2_L(\bold x) \right> = (2 \pi)^{-3} \int d^3 \bold k P_\Phi (k)$} $P_\Phi (k)$:
$
\left < \Phi_L (\bold{k}_1) \Phi_L (\bold{k}_2) \right> = (2 \pi)^3 P_\Phi (k_1) \delta^3 (\bold{k}_1+\bold{k}_2)
$.

  Note that the non-linearity parameter $f_{NL}$  defines 
the non-Gaussianity in the gravitational potential and not in the CMB temperature fluctuations. 

The standard single-field slow-roll inflation model predicts a non-Gaussianity of the form described by Eq.~(\ref{fnl}) \cite{SalopekBond1990,Ganguietal1994,VWHK00,KomatsuSpergel01}.

In standard inflation $f_{NL}$ is unmesurably small (less than $10^{-6}$ \cite{maldacena03, Acquavivaetal03}). Within this picture, the 3-point correlation function (or equivalently, the bispectrum) turns out to be the most sensitive observable to constrain possible departures from these (nearly Gaussian) initial conditions e.g.,\cite{VWHK00}.  Non-linear physics between the end of inflation and the last scattering surface may yield further bispectrum contributions (e.g. sec 8 of  \cite{Bartoloetal04} and references therein and more recent work \cite{ref+1,ref+2, ref+3, BartoloRiotto,Pitrouetal,arXiv-Boubekeur, VerdeMatarrese}) which however are expected to be mostly of equilateral type and below the detection level for forthcoming CMB experiments.  A detection of a non-vanishing CMB bispectrum would be then the smoking gun of a different scenario describing the mechanism responsible for the generation of the primordial density perturbations. 

\subsection{The {\bf Primary} CMB bispectrum}

Here we summarize the equations that describe how the second-order perturbations in the gravitational potential translate into perturbations of the CMB temperature, giving rise to a non-vanishing
contribution to the CMB bispectrum.

For adiabatic scalar perturbations the primary contribution to the CMB coefficients can be written as:
\be\label{primordial-alm}
a_{lm}^P= 4 \pi (-i)^\ell \int \frac{d^3 \bold k }{(2 \pi)^3} \Phi (\bold k) g_{T\ell}(k) Y^*_{\ell m} (\hat{\mathbf n}),
\ee
where $g_{T\ell}(k)$ is the  radiation transfer function and $ \Phi (\bold k)$ is the primordial curvature perturbation in Fourier space. From this equation it is clear that, if any, non-Gaussianity in $ \Phi (\bold k)$ will appear  in the  $a_{lm}^P$.
According to Eq. (\ref{fnl}), we can decompose the curvature perturbation into a linear and non-linear term:
 $\Phi (\bold k) =  \Phi_L (\bold k) +  \Phi_{NL} (\bold k)$
and,  by using an analogous notation, we will have:
$a_{lm}=a_{lm}^L+a_{lm}^{NL}$.

Following the steps outlined in Appendix  \ref{formalism}, the 
primary CMB angular bispectrum takes the form \cite{KomatsuSpergel01}:
\begin{eqnarray}\label{primordial-bisp}
 B_{l_1l_2l_3}^{m_1m_2m_3 \, (P)}
	&=& 2{\cal G}_{l_1l_2l_3}^{m_1m_2m_3}
	\int_0^\infty r^2 dr 
    \left[
          b^{\rm L}_{\ell_1}(r)b^{\rm L}_{\ell_2}(r)b^{\rm NL}_{\ell_3}(r) \right. \nonumber \\
        & +&\left.
b^{\rm L}_{\ell_1}(r)b^{\rm NL}_{\ell_2}(r)b^{\rm L}_{\ell_3}(r) 
	 + b^{\rm NL}_{\ell_1}(r)b^{\rm L}_{\ell_2}(r)b^{\rm L}_{\ell_3}(r)
    \right], \nonumber
\end{eqnarray}
where ${\cal G}_{\ell_1 \ell_2 \ell_3}^{m_1m_2m_3}$ defines the Gaunt integral (see eq. (\ref{gaunt})) and 
\begin{eqnarray}
  \label{eq:bLr}
  b^{\rm L}_{\ell}(r) &\equiv&
  \frac2{\pi}\int_0^\infty k^2 dk P_\Phi(k)g_{{\rm T}\ell}(k)j_l(kr),\\
  \label{eq:bNLr}
  b^{\rm NL}_{\ell}(r) &\equiv&
  \frac2{\pi}\int_0^\infty k^2 dk f_{\rm NL}g_{{\rm T}\ell}(k)j_\ell(kr),
\end{eqnarray}
with $j_\ell(kr)$ being the spherical Bessel functions. 
 It is important to note that this formula is valid only when $f_{NL}$ does not depend on the scale  and it approximately applies if such a dependence is weak. Note that for our present purposes,  if extra contributions are summed to  the primordial one, we can   reinterpret $f_{NL}$  an effective $f_{NL}$ and use the same expression (Eq. \ref{primordial-bisp}) for the primary bispectrum. Extra contributions are not guaranteed to be scale independent or to have exactly the local form,   making the effective $f_{NL}$ shape and scale dependent, as we will see below.
 
 Finally, it is useful to define the
  primary reduced bispectrum factorizing the $f_{NL}$ parameter:
 $
 b^P_{\ell_1 \ell_2 \ell_3}=f_{NL} \,  \hat{b}^P_{\ell_1 \ell_2 \ell_3}$, 
 where the quantity $\hat{b}^{P}_{\ell_1 \ell_2 \ell_3}$ is the reduced bispectrum for $f_{NL}\equiv 1$:
\be\label{red-prim}
\hat{b}^P_{\ell_1 \ell_2 \ell_3}=  B_{\ell_1 \ell_2 \ell_3}^{m_1m_2m_3 \, (P)}\mid_{f_{NL}=1} \, ({\cal G}_{\ell_1 \ell_2 \ell_3}^{m_{1}m_{2}m_{3}})^{-1}\,.
\ee
%
\begin{figure*}
\includegraphics[width=17cm]{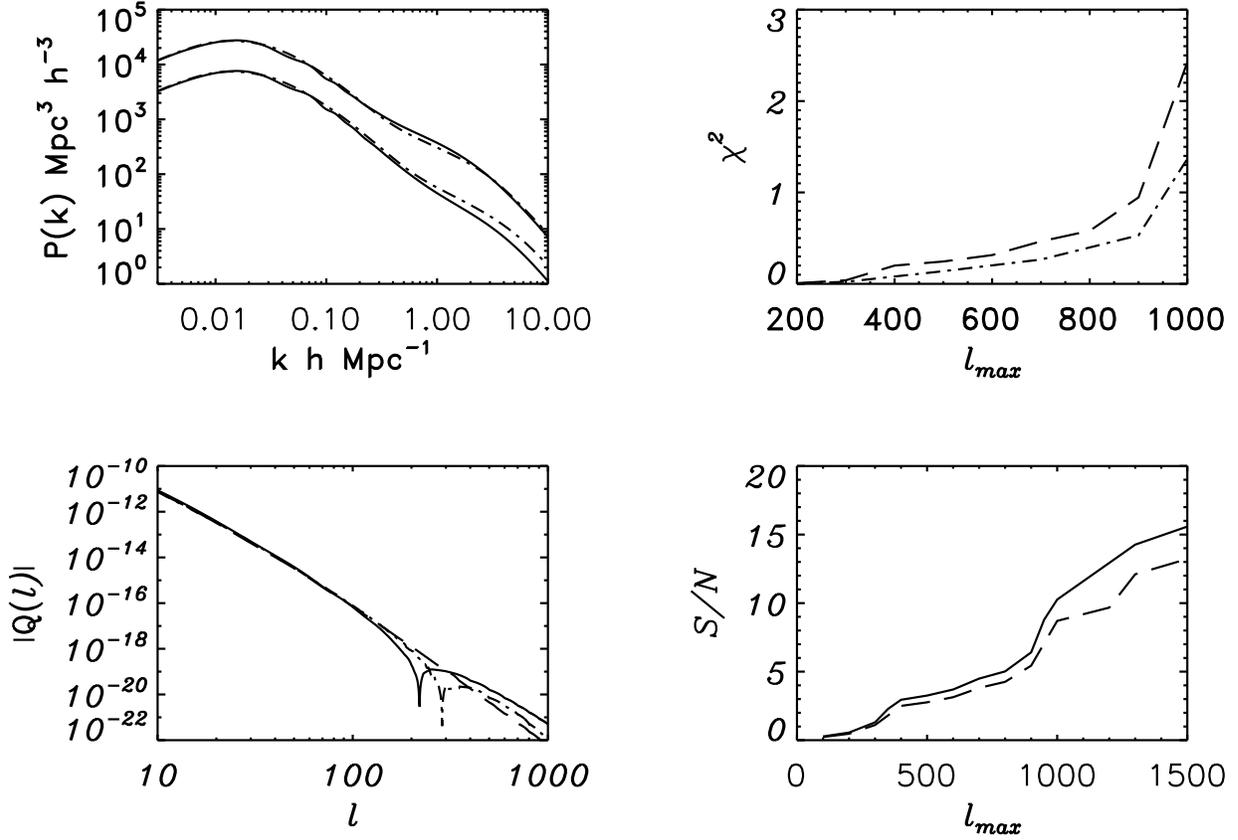}  
\caption{TOP-LEFT panel: The non-linear matter power spectrum $P_\delta^{NL}(k) $ obtained with Halofit (solid line) and by using the Peacock \& Dodds (PD) semi-analytical approach (dot-dashed line). The upper curves refer to redshift $z=0.1$, while the lower curves to $z=1$.
BOTTOM-LEFT panel: The absolute value of the $\cal Q (\ell)$  L-RS bispectrum  coefficients defined in Eq.(\ref{ql}), plotted as a function of the angular scale $\ell$. The cusp indicates where ${\cal Q}$ changes sign due to the onset of non-linearities; in linear theory $\cal Q$ is always positive (dashed line). The solid line corresponds to the coefficients obtained by using the Halofit non-linear matter power spectra $P^{NL}_{\Phi}(k,z)$, while the dot-dashed line refers to the $\cal Q (\ell) $ obtained with the  PD semi-analytical method to model the non-linear behavior. The cosmological parameters used are listed in Tab. \ref{t:param}. Note that the non-linear transition in the two cases happens at different scales: the $\cal Q (\ell)$ from Halofit change sign at $\ell \simeq 210$, while the ones from the PD at $\ell \simeq 300$. 
BOTTOM-RIGHT panel: Signal-to-noise ratio --Eq. (\ref{sn})-- for the secondary Lensing-Rees Sciama Bispectrum as a function of $\ell_{max}$ in the case of an all sky, cosmic variance  limited experiment (solid line). The dashed line refers to the signal-to-noise for the Lensing- linear Integrated Sachs Wolfe bispectrum.
TOP-RIGHT panel: The dot-dashed line is the $\chi^2$ between the L-RS bispectra obtained respectively with Halofit and with PD, as defined in Eq. \ref{chiPD}. The dashed line represents the same quantity, but now the comparison is between the L-RS (Halofit) and the Lensing- linear ISW bispectra. Both quantities are plotted as a function of the maximum multipole $\ell_{max}$. }
\label{fig:ql+pd}
\end{figure*} 

\subsection{Secondary bispectra: the cross correlation between Lensing and the RS effect}\label{L-RS}
The path of the CMB photons traveling from the last scattering surface can be modifed by the gravitational fluctuations along the line-of-sight in several different ways.  On angular scales much larger than arcminute scale the photon's geodesic is deflected by gravitational lensing and late time decay of the gravitational potential  and non-linear growth induce secondary anisotropies known respectively as the Integrated Sachs Wolfe (ISW) \cite{ISW} and the Rees-Sciama (RS) effect \cite{RS}. Hereafter by RS we refer to the combined contribution of linear and non-linear growth.

In this work we will concentrate on the cross correlation of the CMB lensing signal with the secondary anisotropies arising from both the linear ISW and the Rees-Sciama effect. We  will refer to this as L-RS bispectrum. A closely related effect was investigated in \cite{arXiv-Hanson}, where only the (linear)  lSW contribution is included. After galactic foregrounds, point sources and the Sunyaev-Zeldovich \cite{SZ80} signature from galaxy clusters are expected to be the dominant contribution to the CMB bispectrum, but because of their frequency dependence and their statistical properties they can be separated out without major loss of information \cite{KomatsuSpergel01,komatsu-03}.  The next leading  secondary bispectrum contribution is the L-RS one, which cannot be separated out by frequency dependence. Both, lensing and the RS effect, are in fact related to the gravitational potential and thus are correlated, leading to a non-vanishing bispectrum signal with a blackbody spectrum.

As already pointed out in previous works  \cite{licia,Goldberg&Spergel-II, Giovi2003} the joined study of these phenomena through the CMB bispectrum is a very powerful tool, for example,  
 to better understand linear and non-linear growth of structures, 
 to break degeneracies between parameters arising in a power spectrum only analysis,
 or to possibly constrain Dark Energy equation of state or models beyond the standard $\Lambda$CDM. 
 
Following \cite{licia}, the CMB anisotropy in a direction $\hat{n}$ can then be decomposed into:
\be
\Theta(\hat{\mathbf{n}})=\Theta^P(\hat{\mathbf{n}})+\Theta^L(\hat{\mathbf{n}})+\Theta^{RS}(\hat{\mathbf{n}})
\ee
where $P$ denotes
 primary, $L$ lensing (see Eq. \ref{eq:lensingCMB}) and $RS$ ISW+Rees-Sciama, which includes both the linear and the non-linear contributions. This last term takes the form:
\be
\Theta^{RS}(\hat{\mathbf{n}}) = 2 \int dr \frac{\partial }{\partial t} \phi(r,\hat{\mathbf{n}}r),
\ee
where $\phi$ refers to the gravitational potential perturbation and r is the conformal distance defined in Eq. \ref{rz}.

We can thus write the bispectrum (see appendix \ref{formalism}) as
\bea
&&B_{\ell_1 \ell_2 \ell_3}^{m_1 m_2 m_3}\equiv\langle a_{\ell_1}^{m_1}a_{\ell_2}^{m_2}a_{\ell_3}^{m_3}\rangle= \\
&&\langle a_{\ell_1}^{m_1 P}a_{\ell_2}^{m_2 L}a_{\ell_3}^{m_3 RS}\rangle+ 5
\; \mbox{Permutations},.\nonumber
\eea
Following the steps outlined in Appendix \ref{formalism-2} this becomes:
\be \label{bis-l-rs0}
B_{\ell_1 \ell_2 \ell_3}^{m_1 m_2 m_3 \, (L-RS)}={\cal G}^{m_1 m_2m_3}_{\ell_1\ell_2\ell_3}b^{L-RS}_{\ell_1\ell_2\ell_3}
\ee
where the reduced bispectrum is given by 
\bea\label{bis-l-rs}
b_{\ell_1 \ell_2 \ell_3}^{(L-RS)}&= &
\frac{\ell_1(\ell_1+1)-\ell_2(\ell_2+1)+\ell_3(\ell_3+1)}{2} \nonumber\\
 & \times & C_{\ell_1}^P {\cal Q}(\ell_3)
  + 5 \mbox{ Perm.},
\eea
and $C_{\ell}^P$ is the
primary angular CMB power spectrum. 
Here the  quantity that contains  physical information about the late universe is \cite{licia,Goldberg&Spergel-II,cooray-hu}:
\bea\label{ql}
{\cal Q}(\ell)&\equiv& \langle \phi_{L \,\ell}^{*m} a_{\ell}^{RS
m}\rangle \\
&\simeq& 2\int_0^{z_{ls}}\!\!\frac{r(z_{ls})-r(z)}{r(z_{ls})r(z)^3}
\left[
\frac{\partial}{\partial z}P^{NL}_{\phi}(k,z)
\right]_{k=\frac{\ell}{r(z)}}\!\!\!dz \nonumber
\eea
that expresses the statistical expectation of the correlation between the lensing and the RS effect.  Here the Limber approximation has been used, which we find to be extremely good (better than 20\%) even at low $\ell's$. The accuracy of this equation has been explored in \cite{Komatsu2}. The same coefficients can be calculated in the linear case  for the cross correlation lensing-Integrated Sachs-Wolfe effect (ISW) by simply substituting the non-linear power spectrum in equation (\ref{ql}) with the linear one, $P^{L}_{\phi}(k,z)$.

In the bottom-left panel of Fig.~(\ref{fig:ql+pd}) we show the behavior of  the absolute value of these coefficients $|\cal Q (\ell) |$ for $\ell$ up to $1000$ (see \S \ref{numerical} for details). The cusp indicates that   $\cal Q (\ell)$ changes sign: this is due to the onset of non-linearities, which change the sign of $\partial P_{\phi}/\partial z$. Note that in linear theory (dashed line)
such a derivative never changes sign giving \mbox{$\cal Q (\ell) $ $> 0 $} in the $\Lambda$-dominated regime (Integrated Sachs Wolfe effect) \cite{arXiv-Hanson}. Therefore this feature is a fingerprint of the non-linear regime behavior. The scale at which $\cal Q (\ell)$  changes sign depends crucially on the scale at which the non-linear growth overcomes the linear effect, making the the L-RS bispectrum  sensitive to  cosmological parameters governing the growth of structure like $\Omega_m$, $w$ or $\sigma_8$ \cite{licia,Giovi2003,Giovi2005}.

\section{Bispectrum calculation and expected signal-to-noise}\label{numerical}

We assume a fiducial $\Lambda \textrm{CDM}$ model in agreement with the latest observational results \cite{KomatsuWMAP5} with the parameters  listed in Tab. \ref{t:param}.
 The L-RS bispectrum calculation, Eq. (\ref{bis-l-rs}, \ref{ql}), requires evaluation of the non linear $P_{\phi}(k,z)$. The gravitational potential $\phi$ is related to the matter density fluctuation $\delta$  through the Poisson equation:
\be
P_{\phi}(k,z)=\left(\frac{3}{2}\Omega_m\right)^2\left(\frac{H_0}{k}\right)^4 P_\delta(k,z) (1+z)^2\;.
\ee
where $\Omega_m=\Omega_b+\Omega_c$ is the total matter density parameter.
There are two approaches to compute the  $P_\delta^{NL}(k,z)$ that have been extensively tested and used in the literature: the more recent  Halofit \cite{halofit} model, which is included in CAMB \cite{CAMB} and the Peacock and Dodds (PD) \cite{PD} method (generalized for dark energy cosmologies by \cite{maetal99}).
Here we  use both approaches and compare them. Note that the literature so far on the L-RS bispectrum \cite{licia, Giovi2005, Giovi2003} has used the PD approach to describe non-linearities. 

We perform numerical derivatives to map the function $\partial P^{NL}_{\phi}(k,z) / \partial z$ at $k=\frac{\ell}{r(z)}$ with $\ell$ up to $1500$. Then, to compute the $\cal Q (\ell)$  coefficients (see bottom-left panel of Fig.(\ref{fig:ql+pd})), we numerically integrate in $0<z<2.5$, this is sufficient  to account for the dark energy signature and the  non-linear regime (widening the integration interval does not change the results). 
\begin{table} 
\caption{$\Lambda \textrm{CDM}$ parameters.
	}
\begin{center}
\begin{tabular}{ccc} 
\hline
\hline
Symbol&Description&Value\\
\hline
$H_0$&Hubble constant & $70$ Km/sec/Mpc\\
$\Omega_b$&Baryon density&$0.044$\\
$\Omega_c$&Dark matter density&$0.224$\\
$\Omega_\Lambda$&Dark energy density&$0.732$\\
$w$&Dark energy equation of state&$-1$\\
$\sigma_8$&Fluctuation amplitude at $8 h^{-1}$Mpc&$0.834$\\
$n_s$&Scalar spectral index&$1$\\
$z_{ls}$&Redshift of decoupling&$1090.51$\\
\hline
\end{tabular}\label{t:param}
\end{center}
\end{table}
For the 
 primary bispectrum, we proceed as in \cite{KomatsuPhD,KomatsuSpergel01} assuming a $\Lambda$CDM model. We  compute the radiations transfer functions $g_{T\ell}(k)$ with the CMBFAST code \cite{CMBFAST} and we perform the $k$ and $r$-integrations in the same way \cite{KomatsuPhD,KomatsuSpergel01} did.

\subsection{The Lensing-Rees-Sciama bispectrum: Signal-to-Noise ratio}

According to \cite{Goldberg&Spergel-I}, the bispectrum signal-to-noise ratio can be, in general, defined as:
\be\label{sn}
\left( \frac{S}{N}\right)^2 = \sum_{\ell_1 \ell_2 \ell_3} \frac{\left < B_{\ell_1 \ell_2 \ell_3}\right >^2}{\Delta_{\ell_1 \ell_2 \ell_3} \, C_{\ell_1}  C_{\ell_2}   C_{\ell_3}},
\ee
where $\Delta_{\ell_1 \ell_2 \ell_3} $ is a number which takes value $6$ for equilateral configurations, $2$ for isosceles configurations and $1$ otherwise  (see \cite{gangui-b-variance}  for details). 

Using Eq. (\ref{bisp-final}) we can write the the signal in the numerator as:
\bea
 \left < B_{\ell_1 \ell_2 \ell_3} \right>^2
  &=&
  \frac{(2\ell_1+1)(2\ell_2+1)(2\ell_3+1)}{4\pi} \\
 &\times&  \left(
  \begin{array}{ccc}
  \ell_1&\ell_2&\ell_3\\
  0&0&0
  \end{array}
  \right)^2 \, b^{2}_{\ell_1 \ell_2 \ell_3}. \nonumber
  \eea

In the bottom-right panel of Fig. \ref{fig:ql+pd} the solid line shows the signal-to-noise ratio for the L-RS bispectrum as a function of the maximum multipole $\ell_{max}$ ($\ell_1$ , $\ell_2$  and $\ell_3$ are all $< \ell_{max}$). The dashed line refers to the signal-to-noise for the linear case (L-ISW bispectrum). Note the enhancement due to non-linearities at high multipoles. We do not consider $\ell_{max}>1500$ because  other secondary effects (e.g., Ostriker-Vishniac  or Kinetic SZ \cite{OV, JaffeKamionkowski98, SZ80}) may start to dominate. The S/N plotted has been obtained by summing over all triangle configurations for a full sky, ideal,  cosmic variance-dominated experiment.
The results can be representative of an experiment with the nominal performance of Planck, as pointed out in previous works \cite{Giovi2005,Yadav07}.

The signal-to-noise ratio increases mainly when the maximum multipole $\ell_{max}$ reaches few hundred, where the signal gives the main contribution. As we will explore in more detail later on, the L-RS bispectrum signal dominates for squeezed triangle configurations when a large scale mode couples with two small scales modes:  50\% of the signal-to-noise comes from triangles with $2\leq \ell_{min}\leq 10$, in agreement with the findings of \cite{licia}.

\subsection{Modeling non-linearities: Peacock \& Dodds and the Halofit model}\phantom{\footnote{}}

The two main approaches that can be used to compute the non-linear matter power spectrum $P_\delta^{NL}(k)$ are the  commonly used semi-analytical Peacock \& Dodds \cite{PD} (PD) formula, based on the  scaling method of \cite{HKLM} , and the more recent Halofit model \cite{halofit}.
 
  The first approach is based on the {\it ansatz} that the non-linear evolution induce a change of scale so that the non-linear power spectrum  at wavenumber $k$ can be parameterized by  a simple function of the linear one evaluated at $k'$. This has been shown to interpolate correctly the $P(k)$ behavior in the 
 intermediate regime between linear and stable clustering.
  
 The second approach is based on the so-called ``halo model" for the matter power spectrum. In the halo model the density field is decomposed into a distribution of clumps of matter with some density profile. The large scale behavior is then derived through the correlations between different haloes, while the non-linear correlation functions on small scales are obtained from the convolution of the density profile of the halo with itself. Halofit has been also extensively tested on large, high-resolution N-body simulations.

We estimate that any uncertainty in the  description of the non-linear clustering should be at or below the level of the difference between these two approaches.

In the top-left panel of Fig. \ref{fig:ql+pd} the non-linear matter power spectrum $P_\delta^{NL}(k)$ is plotted as a function of  the wavenumber $k$ for Halofit (solid line) and for PD (dot-dashed line). The upper curves refer to power spectra at redshift $z=0.1$, while the lower curves are the non-linear matter power spectra at $z=1$.
The Halofit power spectrum shows the Baryon Acustic Oscillation (BAO) at the typical BAO scale $k\simeq 0.1 \textrm{Mpc} \, $. To produce the PD one we started from a ``no-wiggle" linear power spectrum. This is because the PD approach maps linear scales into non-linear ones thus  artificially changes the position of the wiggles; when taking derivatives this can induce spurious signal which does not happen when staring from a ``no-wiggle" linear $P(k)$.
Beside the BAO feature, which is irrelevant for our purpose,   the two models are in good agreement although at higher $z$ the PD  power spectrum seems to produce a power spectrum more non-linear that Halofit.

The bottom-left panel of Fig. \ref{fig:ql+pd} show the effect of this difference in the L-RS Bispectrum  coefficients $\cal Q (\ell)$. The figure shows the absolute value of the coefficients $| \cal Q (\ell) |$ as a function of the angular scale $\ell$. The solid line corresponds to the coefficients obtained by using Halofit  while the dot-dashed line using PD. The transition to the non-linear regime (indicated by the cusp where  $\cal Q (\ell)$ changes sign) happens at smaller $\ell$ for Halofit   ($\simeq 200$) than for  PD case ($\ell \simeq 300$).

We quantify the difference between the two models by computing the $\chi^2$ for the L-RS bispectra obtained respectively with Halofit and with PD:
\be\label{chiPD}
\chi^2_{Halofit-P\&D}= \sum_{\ell_1 \ell_2 \ell_3} \frac{ \left(B^{L-RS \, [Halofit] }_{\ell_1 \ell_2 \ell_3} - B^{L-RS \, [P\&D]}_{\ell_1 \ell_2 \ell_3}\right)^2}{\Delta_{\ell_1 \ell_2 \ell_3} \, C_{\ell_1}  C_{\ell_2}   C_{\ell_3}}
\ee
 This is shown in the top-right panel of Fig. \ref{fig:ql+pd} (dot-dashed line) where it is plotted as a function of the maximum multipole $\ell_{max}$ for our fiducial cosmology. The two models are compatible within 1-$\sigma$ ($\Delta \chi^2<1$) for $\ell_{max} <~ 900$. 
 We conclude that the significance of a detection of this signal does not depend crucially on the modeling for non-linearities, however the choice of an incorrect modeling may introduce significant biases    when doing precise analysis, as \cite{licia, Giovi2003}, on key parameters, e.g., $w$, $\sigma_8$, $\Omega_m$ which are particularly sensitive to the onset of non-linearity.
 
 For each of the parameters $w$, $\sigma_8$, $\Omega_m$ we compute the bias introduced by using Halofit in the bispectrum calculation in the hypothetical case that PD was a true description of non-linearities. Around our fiducial model, the biases are at the level  comparable to the $1-\sigma$ errors ($0.3$ to $0.6$ $\sigma$).
 We estimate that, in any practical application,  biases introduced by uncertainties in the description of  non-linear clustering will be at this level or below. Moreover, $\partial P^{NL}/\partial z$ can be accurately evaluated with the use of N-body simulations as presented in \cite{Caietal09}, thus removing this source of bias. 
 Ultimately, in the top-right panel of Fig. \ref{fig:ql+pd} we plotted the $\chi^2$ from the L-RS  and the linear Lensing-ISW bispectra. This is defined as in Eq. (\ref{chiPD}), but with the Lensing-ISW bispectrum instead of the non-linear one obtained by using the
PD model. At high multipoles the linear and the non-linear cases
differ by more than 1-$\sigma$. The error introduced by
the change in modeling non-linearities is smaller with respect to the
error due to only consideing the linear behavior.


\section{The Shape-dependence of the Bispectrum  signals: {\bf Primary} vs L-RS}\label{shape}


\begin{figure*}
\centering
\includegraphics[width=17cm]{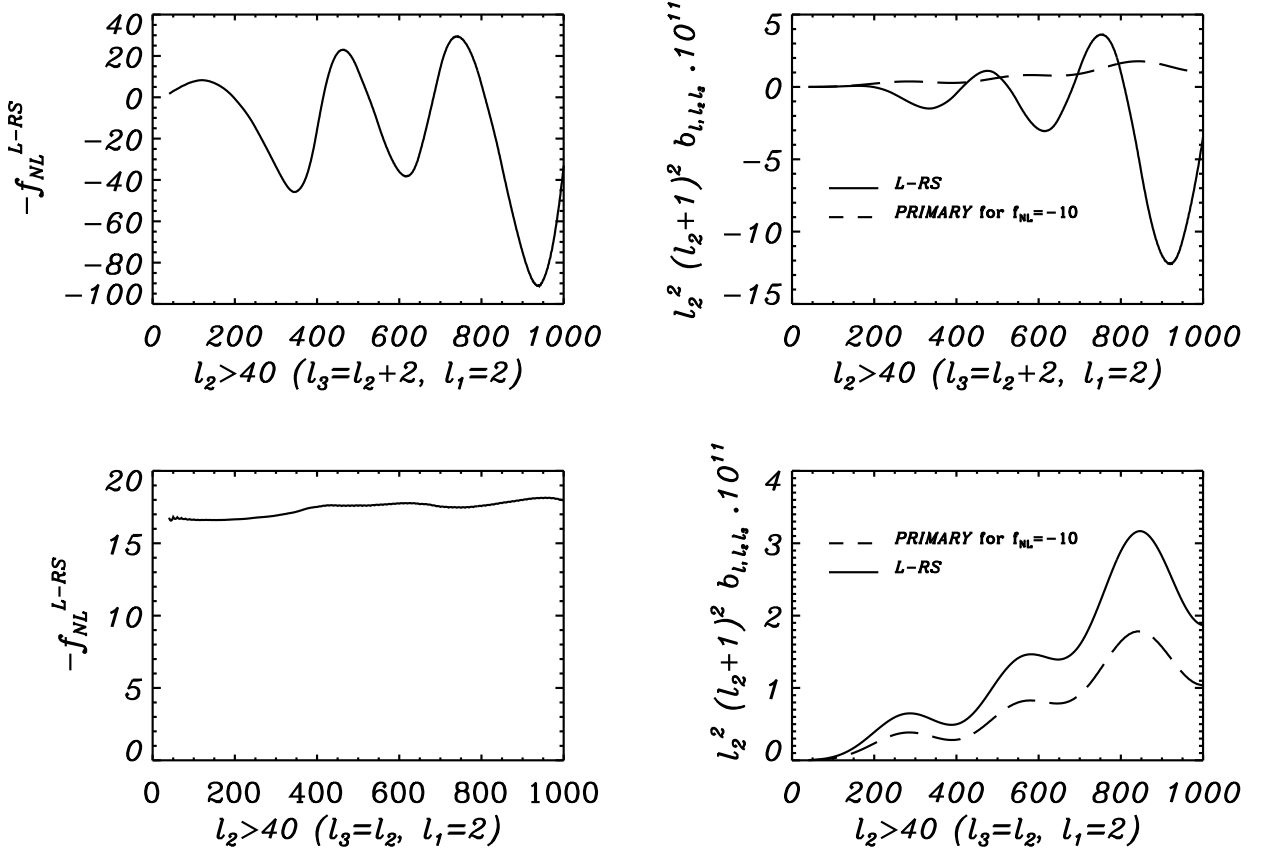} 
\caption{Effective non-linear parameter $f_{NL}^{L-RS}$ (Eq.  \ref{fnl-l-rs}, left panels) and corresponding reduced L-RS (solid) and 
  Primary (Dashed) bispectra (right panels) for two nearly squeezed configurations: $\ell_1=2, \, \ell_2 > 40, \, \ell_3=\ell_2+2$ (top panels) and $\ell_1=2, \, \ell_2 > 40, \, \ell_3=\ell_2$ (bottom panels).  In the right panels the 
  primary bispectrum plotted has $f_{NL}=-10$ for making it more visible.  Note that in the case illustrated in the bottom panels,  the two bispectra have exactly the same shape and they are  completely degenerate for a $f_{NL} \simeq -17$.
}
\label{fig:squeezed}
\end{figure*}

The signal for the
 primary bispectrum is dominated by  squeezed  and nearly squeezed configurations as shown by \cite{creminelli-local,Cabellaetal05,pietrobon09}. We find that the same applies to the L-RS bispectrum, where 90\% of the signal-to-noise  comes form nearly squeezed configurations where $\ell_2> 10 \ell_1$ and $\ell_2<\ell_3$. This can lead to contamination, i.e. ``confusion'', between the two signals.
We illustrate this point by defining an ``effective" $f_{NL}$ for the L-RS signal as
\be\label{fnl-l-rs}
(f^{L-RS}_{NL})_{\ell_1 \ell_2 \ell_3}= \frac{b^{L-RS}_{\ell_1 \ell_2 \ell_3}}{\hat{b}^{P}_{\ell_1 \ell_2 \ell_3}},
\ee
which depends on triangle shape.
This is shown in the left panels of Fig.  \ref{fig:squeezed} while the right panels show the corresponding reduced bispectra  (solid for L-RS and dashed for primary). The top panels are for nearly-squeezed configurations  where $\ell_1$ is fixed, $\ell_1=2$,  $\ell_2$ varies for $\ell_2 > 40$, and  $\ell_3=\ell_2+2$ while the bottom panels are for isosceles squeezed configurations: $\ell_1=2, \, \ell_2 > 40, \, \ell_3=\ell_2$.
 Note that  in the right panels the
 primary bispectrum has been computed for $f_{NL}=-10$  for making it more visible.

The case of the squeezed isosceles configurations (where $\ell_2 \gg \ell_1$, $\ell_3=\ell_2$ and $\ell_1<\sim 150$), contributes with a $\simeq 5\%$ to the total $S/N$ in both cases,  and the two bispectra have exactly the same shape and they completely degenerate for a $f_{NL} \simeq -17$. For nearly squeezed-configurations $f_{NL}^{L-RS}$ oscillates but its average is at around $f_{NL}^{L-RS}\sim 10$. Nearly squeezed configurations where  $\ell_2> 10 \ell_1$ and $\ell_2<\ell_3$ carry most of the S/N. 
\begin{figure*}\label{fig:equilateral}
\centering
\includegraphics[width=17cm]{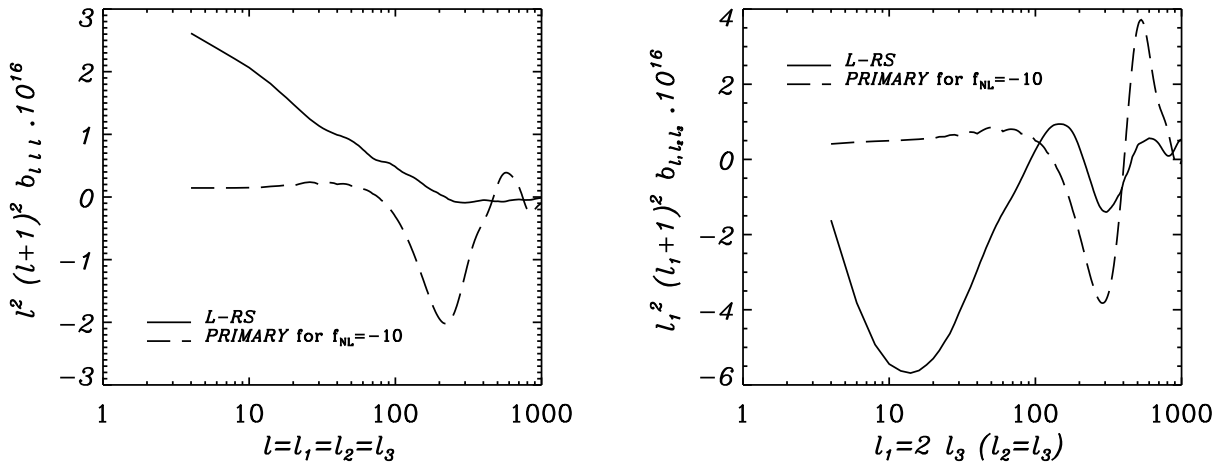}
\caption{Reduced bispectra $b_{l_1l_2l_3}$:
 Primary for $f_{nl}=-10$ (dashed line) and Lensing-Rees Sciama (solid line). 
The left plot shows equilateral triangle configurations $\ell=\ell_1=\ell_2=\ell_3$, while the right one shows the flattened configurations  $\ell_1=2 \ell_3$ and $\ell_2=\ell_3$. 
We have plotted $b^{L-RS}_{l_1l_2l_3}  \ell_1^2 (\ell_1+1)^2 \,10^{16}$, which makes the Sachs-Wolfe plateau of the 
 Primary reduced bispectrum easily seen at large angular scales.}
\label{fig:equilateral}
\end{figure*} 
The L-RS signal always dominates over the 
primary one, normalized to $f_{NL}=1$, for all the high signal-to-noise configurations by a factor of 10-20 in absolute value.

However, besides to the fact that the two bispectra could be confused for showing some similar behavior  (see bottom panels of Fig. \ref{fig:squeezed}) they can in principle be disentangled since they have intrinsically different features arising  from the extremely different physics behind them.

For example, looking back at the top-right panel of Fig. \ref{fig:squeezed}, we find that for these configurations the L-RS signal oscillate, while the primary reduced bispectrum does not. The L-RS bispectrum of eq. (\ref{bis-l-rs}) in fact contains the $C_\ell$, with the typical structure given by the acoustic peaks, and the coefficient $\cal{Q}_\ell$ which determine the change of sign.
On the other hand, the
 primary signal, see eq. (\ref{primordial-bisp}), is composed by the coefficients: $b_\ell^L(r) \varpropto P_\phi(k) g_{T\ell}(k)$ and 
$b_\ell^{NL}(r) \varpropto f_{NL} g_{T\ell}(k)$ so that the changing of sign in this case is due to the full radiation transfer functions $g_{T\ell}(k)$.
For general configurations, the two bispectra behave differently: in Fig. \ref{fig:equilateral} we plot the case of equilateral (left panel) and flattened configurations of the type: $\ell_1=2  \ell_3$ and $\ell_2=\ell_3$ (right panel). The dashed lines refers to the 
primary contribution while the solid lines to the L-RS one. The two bispectra  have different shapes and change sign at different angular scales. In the case of equilateral configurations, for example, the  primary reduced bispectrum shows the known oscillatory shape, as found in \cite{KomatsuSpergel01}, while the L-RS reduced bispectrum does not.  In the case of flattened configurations both bispectra show oscillations.
Note that in these plots the y-axis has been multiplied by a factor $10^{16}$ (while in Fig.~(\ref{fig:squeezed}) the y-axis has been multiplied by a factor $10^{11}$): these contributions are clearly subdominant by about 5 orders of magnitude with respect to the squeezed configurations, which explains why the latter shapes dominate the  signal-to-noise. 

In light of these findings we now attempt to interpret recent constraints on primordial non-Gaussianity from CMB data e.g., \cite{Yadav08, KomatsuWMAP5,CreminelliWMAP3} and consider the  implications for forthcoming measurements.

The $f_{NL}$ estimator used in these works reduces to the one defined in \cite{KomatsuSpergelWandelt05} in the simplest case of temperature-only anisotropies, cosmic variance dominated, all sky analysis. This estmator weights the bispectrum of every triplet $\ell_1,\ell_2,\ell_3$  by the signal to noise of the
 primary bispectrum.  We can thus estimate the 
contamination that such an estimator would measure due to the presence of the L-RS signal defining:
\be\label{eq:fnl-est}
\hat{f}_{NL}= \frac{\hat{S}}{N},
\ee
where
\be
\hat{S}= \sum_{2 \leqslant \ell_1 \ell_2 \ell_3} \frac{ B^{L-RS}_{\ell_1 \ell_2 \ell_3} \, B^{P}_{\ell_1 \ell_2 \ell_3} }{ C_{\ell_1}  C_{\ell_2}   C_{\ell_3} }
\ee
and
\be
N= \sum_{2 \leqslant \ell_1 \ell_2 \ell_3} \frac{ (B^{P}_{\ell_1 \ell_2 \ell_3})^2 }{ C_{\ell_1}  C_{\ell_2}   C_{\ell_3} }.
\ee

This is plotted as a function of $\ell_{max}$ in Fig. \ref{fig:fnl-est}, up to $\ell_{max}=1500$: the dashed line refers to $\hat{f}_{NL}$ obtained by summing over all configurations, while the dot-dashed line refers to $\hat{f}_{NL}$ obtained from only  nearly squeezed configurations ($\sum_{\ell_1=2}^{10}\sum_{\ell_2=50 \ell_1}^{\ell_{max}}\sum_{\ell_3=\ell_2}^{\ell_{max}}$), which dominate for both the primary (local type) and the lensing-Rees Sciama bispectrum. The solid lines indicates where the bias is negative. This is in qualitative agreement with the effect explored in \cite{arXiv-Hanson} where only linear growth was included: the inclusion of (later-type) non-linearities enhances the contamination but, as expected, only for $\ell >400$. For example at $\ell \sim 1500$ we estimate an enhancement of roughly 40\%.

Using Halofit or PD in the modeling of the L-RS bispectrum changes the estimates of the effective $f_{NL}$ by 10\% indicating that this correction is robust to possible residual uncertainty in the modeling of non-linearities.

It is also possible to estimate the effective $f_{NL}$ via a simple $\chi^2$ analysis: we find the same values as above but the interpretation of the $\chi^2$ as a goodness of fit test would indicate that the local model is not a good fit.  This is in qualitative agreement with the findings of \cite{Nittaetal09} who  compute the CMB bispectrum from the second-order fluctuations and find that their effect is separable from the primary non-Gaussian signal because of the different shape dependence for non-squeezed (or nearly squeezed) configurations. The agreement cannot be made fully quantitative as perturbation theory approach may break down: for a given multipole $\ell$ the derivative of the  gravitational potential power spectrum is probed at a wide range of scales $k(z)=\frac{\ell}{r(z)}$ and therefore  highly non-linear scales can contribute non-negligibly even at relatively low $\ell$. In practice, however, it may not always be possible to implement a goodness of fit test.

The expected error on $f_{NL}$ for forthcoming surveys is smaller than $10$ (for example the Planck surveyor, recently launched is expected to yield 1-$\sigma$ error on $f_{NL}$ of order 4 \cite{Yadav07}), indicating that the L-RS signal may be a crucial contaminant in the pursuit of primordial non-Gaussianity, if not properly taken into account.  We have shown here that its amplitude and configuration dependence is well known; it is thus not necessary to extract this signal from the CMB bispectrum and separate it from the primary:  it can simply be included in the  modeling of the CMB bispectrum.

 \section{conclusions}\label{end}
 We have revisited the predictions for the expected CMB bispectrum signature of the
primary-lensing Rees-Sciama (L-RS) correlation. This bispectrum is the leading secondary  contribution on scales much larger than arcminute with the same frequency dependence as the CMB primary. Forthcoming experiments  like Planck have the statistical power to detect this signal with a signal-to-noise of order 10.  
 The linear contribution (primary-
 lensing-ISW bispectrum) was considered in \cite{Goldberg&Spergel-II, CoorayHu2001} and in \cite{arXiv-Hanson}. By including the non-linear (RS) description  the signal-to-noise increases to $\simeq 10$ to $\ell_{max}=1000$.  
The overall  signal depends on  the balance of two competing contributions along the line 
of sight: the decaying gravitational potential fluctuations and the amplification due to non-linear gravity.  For this reason the effect can be used to place strong constraints on cosmological parameters that determine the growth of structures: $\Omega_m$, dark energy parameters and $\sigma_8$. By comparing two different semi-analitic descriptions of non-linear clustering, we find that an accurate description of the non-linear growth of the matter power spectrum is necessary to obtain unbiased estimates of these parameters.  Approaches based on numerical simulations (see e.g., \cite{Komatsu2,Caietal09}) will have to be employed. In general, while the approximations used here to  derive and compute  Eq.\ref{ql} are extremely good for the purpose of this paper, a detailed comparison with data will require the exact numerical evaluations.

\begin{figure}\label{fig:fnl-est}
\includegraphics[width=8.5cm]{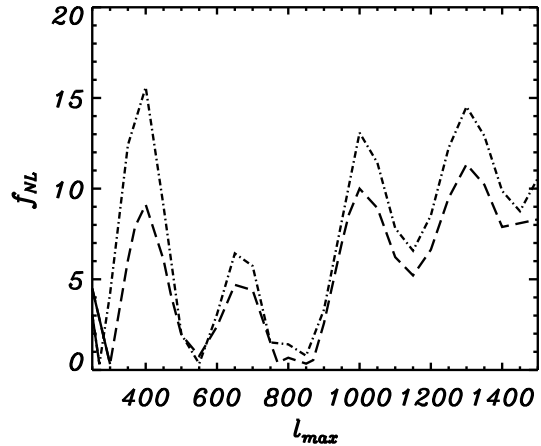}
\caption{The plot shows $\hat{f}_{NL}$, as defined in Eq. \ref{eq:fnl-est}, as a function of $\ell_{max}$. The dashed line refers to $\hat{f}_{NL}$ obtained by summing over all configurations, while the dot-dashed line refers to $\hat{f}_{NL}$ obtained from nearly squeezed configuration (where $\ell_1$ runs from 2 to 10, $\ell_2$ from 50$\ell_1$ to $\ell_{max}$ and $\ell_3$ from $\ell_2$ to $\ell_{max}$), which dominate for both the primary (local type) and the lensing-Rees Sciama bispectrum. The solid lines indicates where the bias is negative.
}
\label{fig:fnl-est}
\end{figure}

Here we have shown that  this bispectrum signal can be confused with the signal from local primordial non-Gaussianity. Both bispectra signal are maximal for squeezed or nearly squeezed configurations. For some configurations (e.g., squeezed isosceles) the  two bispectra  are virtually identical, while  for generic configurations the   shape dependence of the two bispectra are different in the details. A bispectrum  estimator optimized for constraining primordial non-Gaussianity of the local type would measure an effective $f_{NL}=10$ for $\ell_{max}=1000$ due to the presence of the  primary-
lensing-Rees-Sciama correlation.
If not accounted for, this introduces a contamination in the constraints on primordial non-Gaussianity from the CMB bispectrum. 
This is in qualitative agreement with the effect explored in \cite{arXiv-Hanson} where only linear growth was included. For $\ell>400$ the full non-linear treatment is needed. For current data, this  contamination (effectively  bias  in the recovered $f_{NL}$) is smaller than the 1-$\sigma$ error, however it can become significant when interpreting the statistical significance of results that are at the boundary of the $3-\sigma$ confidence level. For example if  we subtract the effective value for the L-RS $f_{NL}$ from the central value of the estimate of \cite{Yadav08}, we obtain that $f_{NL}$ primordial is consistent to zero at the  $\sim 2.5 \sigma$ confidence level. For forthcoming data, however, this bias will be larger than the $1-\sigma$ error and thus non-negligible. A more quantitative statement cannot be made  at this stage because the calculations presented here are done for a cosmic-variance dominated experiments while  for the current bispectrum analysis from WMAP data instrumental noise cannot be neglected at $\ell \sim 400$ where most of the contamination is expected to come from.

Techniques to  separate out different bispectra shapes and   assess whether a detection 
of non-Gaussianity is primordial have been proposed \cite{MunshiHeavens09} and will be suitable for this application.

We argue that the bispectrum of the L-RS  effect can be accurately modeled: even with currently available semi-analytic descriptions for non-linear clustering,  we estimate the error on the  effective $f_{NL}$  to be at  the 10\% level or below. 
We conclude that, in analyzing the CMB bispectrum to obtain constraints on primordial non-Gaussianity for forthcoming data, this contribution must be included in the modeling.

 
  
 \section*{Acknowledgments}
 We thank Michele Liguori for invaluable help with the CMBFAST  radiation transfer function, Eiichiro Komatsu for insightful comments on an earlier version of the manuscript and Alan Heavens, Benjamin Wandelt and Sabino Matarrese for useful feedback. 
 AM is supported by  CSIC I3 \#200750I034 and FP7-PEOPLE-2002IRG4-4-IRG\#202182. LV acknowledges support of FP7-PEOPLE-2002IRG4-4-IRG\#202182.
 
 \addcontentsline{toc}{chapter}{Appendix}

\appendix

\section{Bispectrum statistics}\label{formalism}
Deviations form Gaussianity in the CMB are characterized by the angular n-points correlation function of the temperature field in the sky \cite{luo-94}:
\be
\left<\Theta(\hat{\mathbf{n}}_1) \Theta(\hat{\mathbf{n}}_2) ...  \Theta(\hat{\mathbf{n}}_n) \right> 
\ee
where the bracket defines the ensemble average and $\hat{\mathbf{n}}$ the angular position (i.e. the direction unit vector of the incoming photons).
 In general it is useful to expand the field in terms of spherical harmonics: 
\be
\Theta(\hat{\mathbf{n}})=\sum_{\ell=0}^\infty \sum_{m=- \ell}^{\ell} a_{\ell m} Y_{\ell m}(\hat{\mathbf{n}}),
\ee
so that, by using the symmetric proprieties of harmonic transformations and the orthogonality of the spherical harmonics, we can write the coefficients $a_\ell^m$ as:
\begin{equation}\label{alm}
a_{\ell}^m=\int \; d^2\hat{n} \;  \Theta (\hat{\mathbf{n}}) Y^{*m}_\ell(\hat{\mathbf{n}}). 
\ee

The angular CMB bispectrum is defined by three harmonic transforms satisfying rotational invariance:
 \be
 B_{\ell_1 \ell_2 \ell_3}^{m_1m_2m_3}\equiv 
  \left<a_{\ell_1m_1}a_{\ell_2m_2}a_{\ell_3m_3}\right>\,,
 \ee
thus  the angular averaged bispectrum takes the form: 
 \begin{equation}\label{bispectrum}
  B_{\ell_1 \ell_2 \ell_3}= \sum_{{\rm all}~m}
  \left(
  \begin{array}{ccc}
  \ell_1&\ell_2&\ell_3\\
  m_1&m_2&m_3
  \end{array}
  \right)
  B_{\ell_1 \ell_2 \ell_3}^{m_1m_2m_3}.
\end{equation}
Since $\ell_1$,$\ell_2$ and $\ell_2$ form a triangle, this quantity must satisfy the triangle conditions and parity invariance:
\be
m_1+m_2+m_3=0, \;  \; \; \ell_1+\ell_2+\ell_3={\rm even},  \nonumber
\ee
\begin{center}
\be
\left|\ell_i-\ell_j\right|\leq \ell_k \leq \ell_i+\ell_j
\ee
\end{center}
 for all permutations of indices. 
The matrix appearing in equation (\ref{bispectrum}) represents the Wigner-3j symbol  that describes the coupling of two angular momenta. Rotational invariance requires the bispectrum amplitude to be independent from orientation and triangle configuration. The Wigner-3j symbol, transforming the m's under rotations, preserve the triangle configuration thus describing the bispectrum azimuthal angle dependence.
The orthogonality properties of the  Wigner-3j symbols are
\be
\sum_{{\rm all}~m}  \left(
  \begin{array}{ccc}
  \ell_1&\ell_2&\ell_3\\
  m_1&m_2&m_3
  \end{array}
  \right)^2=1
\ee
\be
\sum_{m'_1\;m'_2}  \left(
  \begin{array}{ccc}
  \ell_1&\ell_2&\ell_3\\
  m'_1&m'_2&m'_3
  \end{array}
  \right)\;
   \left(
  \begin{array}{ccc}
  \ell_1&\ell_2&L\\
  m'_1&m'_2&M'
  \end{array}
  \right)= \frac{\delta_{\ell_3 L}\delta_{m'_3 M'}}{2 L +1}
\ee
 By making use again of rotational invariance and of the simmetry and orto-normality properties of the 3-j symbols, we can write the bispectrum as:
 \be\label{reduced-bis}
 B_{\ell_1 \ell_2 \ell_3}^{m_1m_2m_3}={\cal G}^{m_1m_2 m_3}_{\ell_1 \ell_2 \ell_3} b_{\ell_1 \ell_2 \ell_3}
\ee
where ${\cal G}^{m_1m_2 m_3}_{\ell_1 \ell_2 \ell_3}$ is the Gaunt integral which contains all the angle dependence and triangle constraint information and it is defined by:
\bea\label{gaunt}
  {\cal G}_{\ell_1 \ell_2 \ell_3}^{m_1m_2m_3}
  &\equiv&
  \int d^2\hat{\mathbf n}
  Y_{\ell_1m_1}(\hat{\mathbf n})
  Y_{\ell_2m_2}(\hat{\mathbf n})
  Y_{\ell_3m_3}(\hat{\mathbf n}) \nonumber \\
  \label{eq:gaunt}
  &=&\sqrt{
   \frac{\left(2\ell_1+1\right)\left(2\ell_2+1\right)\left(2\ell_3+1\right)}
        {4\pi}
        }\\
  &\times&\left(
  \begin{array}{ccc}
  \ell_1 & \ell_2 & \ell_3 \\ 0 & 0 & 0 
  \end{array}
  \right)
  \left(
  \begin{array}{ccc}
  \ell_1 & \ell_2 & \ell_3 \\ m_1 & m_2 & m_3 
  \end{array}
  \right),\nonumber
\end{eqnarray}
where $b_{\ell_1 \ell_2 \ell_3}$ is called the reduced bispectrum, which is a very useful quantity since it is an arbitrary symmetric function of $\ell_1$,$\ell_2$ and $\ell_3$ only and it contains all the relevant physical information of the bispectrum.  

By substituting equation(\ref{reduced-bis}) into equation(\ref{bispectrum}) and using the Gaunt integral property:
\bea
  \label{eq:wigner}
  \sum_{{\rm all}~m} &&
  \left(
  \begin{array}{ccc}
  \ell_1&\ell_2&\ell_3\\
  m_1&m_2&m_3
  \end{array}
  \right)
  {\cal G}_{\ell_1 \ell_2 \ell_3}^{m_1m_2m_3}=\\
 &=& 
  \sqrt{\frac{(2\ell_1+1)(2\ell_2+1)(2\ell_3+1)}{4\pi}}
  \left(
  \begin{array}{ccc}
  \ell_1&\ell_2&\ell_3\\
  0&0&0
  \end{array}
  \right), \nonumber
\eea
we can finally write:
\bea
  \label{bisp-final}
  B_{\ell_1 \ell_2 \ell_3}
 & =&
  \sqrt{\frac{(2\ell_1+1)(2\ell_2+1)(2\ell_3+1)}{4\pi}} \\
  &\times& \left(
  \begin{array}{ccc}
  \ell_1&\ell_2&\ell_3\\
  0&0&0
  \end{array}
  \right)b_{\ell_1 \ell_2 \ell_3}. \nonumber
\eea

For high-$\ell$ the  Gosper factorials approximation for the Wigner 3j symbols can be used:
 \bea\label{Gosper}
&&  \left( 
\begin{array}{ccc}
\ell_{1} & \ell_{2} & \ell_{3} \\ 
0 & 0 & 0
\end{array}
\right) \simeq \left( -\frac{L}{L+1}\right) ^{L/2}\frac{1}{\left( 6L+7\right)^
{1/4}} \\
&& \left( \frac{3e}{\pi }\frac{3L+1}{L+1}\right)^{1/2}\prod\limits_{i=1}^{3}
\frac{\left( 6L-12\ell_{i}+1\right) ^{1/4}}{\left( 3L-6\ell_{i}+1\right) ^{1/2}}\ .\nonumber
  \eea
%
%

\section{Weak lensing of the CMB}\label{formalism-2}


Weak lensing of the CMB re-maps the temperature primary anisotropy  according to:
\bea\label{eq:lensingCMB}
\Theta^L(\hat{\mathbf{n}})&=& \Theta^P(\hat{\mathbf{n}} + \nabla \phi) \nonumber\\
&\simeq &  \Theta^P(\hat{\mathbf{n}} )+ \nabla_i \phi(\hat{\mathbf{n}}) \nabla^i  \Theta^P(\hat{\mathbf{n}} ) + \ldots 
\eea

where the label 'L' refers to the lensed term while 'P' to the primary contribution. The deflection angle $\mathbf{\alpha}= \nabla \phi_L$ is given by the angular gradient of the gravitational potential projection along the line of sight:
\be
\phi_L(\hat{\mathbf{n}})= -2 \int^{r_{ls}}_0 dr \frac{r(z_{ls})-r(z)}{r(z) \, r(z_{ls}) } \, \phi(r, \hat n r).
\ee

Here $r$ is the comoving conformal distance. 
Assuming a flat $\Lambda$CDM universe this can be written as:
\begin{equation}\label{rz}
r \left( z \right) = 
\frac{c}{H_0}\int_0^z\frac{dz'}{\sqrt{\Omega_{m0} \left( 1 + z^{\prime} \right)^{3} + \Omega_{\Lambda  0}}}.
\end{equation}

and thus $r_{ls}\equiv r(z_{ls})$ refers to the comoving radius at last scattering from the observer at $z=0$.

As done with the temperature perturbations, we can expand the lensing potential into multipole moments:
\be
 \phi_L(\hat{\mathbf{n}})= \sum_{\ell m} \phi_{L \, \ell}^{m} Y_{\ell}^{m} (\hat{\mathbf{n}})\,.
\ee

By applying eq. (\ref{alm}) into eq. (\ref{eq:lensingCMB}) and carrying out the calculations we get an explicit expression for the lensing $a_l^m$ coefficients:
\bea\label{lensing-alm}
a_{\ell}^{m L}& = & a_{\ell}^{mP}+\sum_{\ell'\ell''m' m''}(-1)^{m+m'+m''}{\cal G}^{-m m' m''}_{\ell \ell' \ell''} \\
& \times &\frac{\ell'(\ell'+1)-\ell(\ell+1)+\ell''(\ell''+1)}{2}a^{m'P*}_{\ell'}\phi^{*
-m''}_{L \, \ell''}\nonumber
\eea

being ${\cal G}^{m m' m''}_{\ell \ell' \ell''} $ the Gaunt integral defined in equation(\ref{gaunt}).

\end{document}